\documentclass{mn2e}
\usepackage{psfig}

\catcode`\@=11
\def\gsim{\ifmmode{\mathrel{\mathpalette\@versim>}}
    \else{$\mathrel{\mathpalette\@versim>}$}\fi}
\def\lsim{\ifmmode{\mathrel{\mathpalette\@versim<}}
    \else{$\mathrel{\mathpalette\@versim<}$}\fi}
\def\@versim#1#2{\lower 2.9truept \vbox{\baselineskip 0pt \lineskip
    0.5truept \ialign{$\m@th#1\hfil##\hfil$\crcr#2\crcr\sim\crcr}}}
\catcode`\@=12

\newcommand{\beq}{\begin{equation}}
\newcommand{\eeq}{\end{equation}}
\newcommand{\az}{{a_0}}
\newcommand{\phiN}{\phi^{\rm N}}
\newcommand{\gv}{{\bf g}}
\newcommand{\gvN}{{\bf g}^{\rm N}}
\newcommand{\Sv}{{\bf S}}
\newcommand{\rhos}{\rho_*}
\newcommand{\Mstar}{M_*}
\newcommand{\Mstarten}{M_{*,10}}
\newcommand{\tstar}{t_*}
\newcommand{\MDM}{M_{\rm DM}}
\newcommand{\rs}{r_*}
\newcommand{\vstar}{v_*}

\newcommand{\rhalf}{r_{\rm M}
}\newcommand{\rhalfz}{r_{\rm M,0}}
\newcommand{\dz}{d_{\rm 0}}
\newcommand{\bz}{b_{\rm 0}}
\newcommand{\vz}{v_{\rm 0}}
\newcommand{\vrel}{v_{\rm rel}}
\newcommand{\drest}{d_{\rm rest}}
\newcommand{\Nstar}{N_{*}}
\newcommand{\NDM}{N_{\rm DM}}
\newcommand{\sigmavz}{\sigma_{\rm V,0}}
\newcommand{\sigmav}{\sigma_{\rm V}}

\newcommand{\rhodm}{\rho_{\rm DM}}

\newcommand{\bey}{\begin{eqnarray}}
\newcommand{\eey}{\end{eqnarray}}
\newcommand{\Myr}{\, {\rm Myr} }
\newcommand{\Gyr}{\, {\rm Gyr} }

\newcommand{\kpc}{\, {\rm kpc} }
\newcommand{\Msun}{M_\odot}

\newcommand{\kms}{\, {\rm km \, s}^{-1} }

\newcommand{\xv}{{\bf x}}

\def\av#1{\langle#1\rangle}
\def\mav{\av{m}}

\def\Re{R_{\rm e}}

\def\xv{{\bf x}}

\def\xv{{\bf x}}

   \title[Galaxy merging in MOND]{Galaxy merging in MOND}

   \author[Nipoti et al.]
          {Carlo Nipoti$^1$, Pasquale Londrillo$^2$, and 
           Luca Ciotti$^1$
           \\ $^1$Astronomy Department, University of Bologna, 
                       via Ranzani 1, 40127 Bologna, Italy
           \\ $^2$INAF-Bologna Astronomical Observatory, 
                       via Ranzani 1, 40127 Bologna, Italy
           }

\date{Accepted 2007 August 7.  Received 2007 July 25; in original form 2007 May 31}
\pubyear{2007}

\begin{document} 
\maketitle

\begin{abstract} 
  We present the results of N-body simulations of dissipationless
  galaxy merging in Modified Newtonian Dynamics (MOND). For
  comparison, we also studied Newtonian merging between galaxies
  embedded in dark matter halos, with internal dynamics equivalent to
  the MOND systems.  We found that the merging timescales are
  significantly longer in MOND than in Newtonian gravity with dark
  matter, suggesting that observational evidence of rapid merging
  could be difficult to explain in MOND. However, when two galaxies
  eventually merge, the MOND merging end-product is hardly
  distinguishable from the final stellar distribution of an equivalent
  Newtonian merger with dark matter.
\end{abstract}


\section{Introduction}
\label{secint}

Given the remarkable ability of MOND to reproduce the kinematics of
galaxies (e.g., Milgrom~2002; Sanders \& McGaugh~2002) and its
increased interest due to the possibility of a relativistic
formulation (Bekenstein~2004), it is natural to look for tests able to
discriminate between MOND and Newtonian gravity with dark matter (DM).
Despite numerous attempts, no clear-cut cases have been found so far
(see Bekenstein~2006), the main reason being that MOND is a non-linear
theory, and this makes the study of systems deviating significantly
from spherical symmetry (Brada \& Milgrom~1995; Ciotti, Londrillo \&
Nipoti~2006) and the performance of N-body simulations more difficult
in MOND than in Newtonian dynamics. In particular, very few N-body
simulations of MOND systems have been performed so far, exploring
stability of disk galaxies (Brada \& Milgrom~1999; Tiret \&
Combes~2007), external field effect (Brada \& Milgrom~2000),
dissipationless collapse and phase mixing (Nipoti, Londrillo \& Ciotti
2007a, hereafter NLC07; Ciotti, Nipoti \& Londrillo~2007). In
addition, the study of structure formation in MOND using N-body
cosmological simulations is still limited to a couple of preliminary
explorations (Nusser~2002; Knebe \& Gibson~2004).

Observations leave no doubt that galaxy merging occurs (e.g. Arp~1966;
Schweizer~1982), and it is also known that Newtonian gravity can
account in detail for such a process (Toomre \& Toomre~1972).  It is
then natural to study galaxy merging in MOND.  In fact, there are
reasons to expect that galaxy merging is less effective in MOND than
in Newtonian gravity: in MOND galaxies are expected to collide at high
speed, and there are no DM halos to absorb orbital energy and
angular momentum (Binney~2004; Sellwood~2004); in addition, it has
been recently shown that violent relaxation and phase mixing are
slower in MOND (NLC07; Ciotti et al.~2007).

Taking advantage of our recently developed MOND N-body code, in this
Letter we present the results of N-body simulations of galaxy merging
in MOND, focusing for simplicity on the case of dissipationless
merging between equal-mass spherical galaxies.  For comparison, we
also consider the merging of structurally identical purely baryonic
Newtonian systems, and the merging of equivalent Newtonian systems
(i.e., Newtonian models with the same baryonic distribution as the
MOND systems, embedded in DM halos such that their internal dynamics
matches the corresponding MOND cases; see Milgrom 2001; Nipoti et
al. 2007b).

\begin{table*}

 \flushleft{


  \caption{Parameters of the simulations and properties of the end-products.}

  \begin{tabular}{llcccccccccccc}
     Name    & Gravity &  $\MDM/\Mstar$ & $\kappa$ & $\vz/\vstar$ &      $\bz/\dz$ & $\Nstar$ &  $\NDM$ & $c/a$ & $b/a$ &         $\rhalf/\rhalfz$ & $\sigmav/\sigmavz$& $\gamma$ & $\mav$\\
~~(1) & ~~~(2) & (3) & (4) & (5) &  (6) & (7) & (8) & (9) & (10) & (11) & (12) & (13) & (14) \\ 
\hline
M1h   & MOND   &~0 & ~1  & 0.958 &0  & $2\times10^6$ & 0           &0.52&0.54&1.6&1.09&$1.4\pm0.3$&$4.4\pm0.3$\\
E1h   & Newton &30 & ~1  & 1.017 &0  & $2\times10^5$ &$2\times10^6$&0.48&0.50&1.2&1.22&$1.0\pm0.2$&$3.4\pm0.3$\\
M1o   & MOND   &~0 & ~1  & 0.958 &0.5& $2\times10^6$ & 0           &0.51&0.61&1.9&1.09&$2.0\pm0.3$&$6.3\pm0.3$\\
E1o   & Newton &30 & ~1  & 1.017 &0.5& $2\times10^5$ &$2\times10^6$&0.54&0.60&1.2&1.23&$1.3\pm0.2$&$4.1\pm0.2$\\
M25h  & MOND   &~0 & 25  & 0.428 &0  & $2\times10^6$ & 0           &0.54&0.73&2.1&1.16&$1.8\pm0.3$&$4.2\pm0.3$\\
E25h  & Newton &~5 & 25  & 0.447 &0  & $2\times10^5$ &$1\times10^6$&0.52&0.57&1.5&1.27&$1.3\pm0.3$&$3.5\pm0.3$\\
M25o  & MOND   &~0 & 25  & 0.428 &0.5& $2\times10^6$ & 0           &0.59&0.79&2.2&1.16&$2.0\pm0.3$&$6.3\pm0.5$\\
E25o  & Newton &~5 & 25  & 0.447 &0.5& $2\times10^5$ &$1\times10^6$&0.62&0.83&1.3&1.29&$1.7\pm0.3$&$4.5\pm0.3$\\
N0h   & Newton &~0 & ~-  & 0.183 &0  & $2\times10^6$ & 0           &0.68&0.71&2.0&1.07&$2.0\pm0.1$&$5.0\pm0.3$\\
N0o   & Newton &~0 & ~-  & 0.183 &0.5& $2\times10^6$ & 0           &0.64&0.89&1.6&1.07&$2.1\pm0.2$&$5.5\pm0.3$\\
\hline

\end{tabular}

}

\medskip

\flushleft{(1): name of the simulation. (2): gravity law. (3): DM to
baryonic mass ratio. (4): internal acceleration ratio. (5): normalised
relative speed of the barycentres at $t=0$. (6): normalised impact
parameter at $t=0$ (in all cases $\dz=40\rs$).  (7): total number of
stellar particles. (8): total number of DM particles. (9): end-product
minor-to-major axis ratio. (10): end-product intermediate-to-major
axis ratios. (11): final-to-initial half-mass radius ratio.  (12):
final-to-initial virial velocity dispersion ratio. (13): best-fit
inner logarithmic slope of the final angle-averaged density
profile. (14): best-fit Sersic index of the final projected density
profile.}
\end{table*}

\section{The numerical simulations}
\label{secmet}

We consider MOND in Bekenstein \& Milgrom's (1984) formulation, in
which the Poisson equation $\nabla^2\phiN=4\pi G\rho$ is substituted
by the non-relativistic field equation
\begin{equation}
\nabla\cdot\left[\mu\left({\Vert\nabla\phi\Vert\over\az}\right)
\nabla\phi\right] = 4\pi G \rho.
\label{eqMOND}
\end{equation} In the equation above $\Vert ...\Vert$ is the standard Euclidean
norm, $\phi$ and $\phiN$ are, respectively, the MOND and Newtonian
gravitational potentials produced by $\rho$, and for finite mass
systems $\nabla\phi\to 0$ for $\Vert\xv\Vert\to\infty$.  The function
$\mu(y)$ is not constrained by the theory except that it must run
smoothly from $\mu(y)\sim y$ at $y\ll 1$ (in the so-called deep-MOND
regime) to $\mu(y)\sim 1$ at $y\gg 1$, with a dividing acceleration
scale $\az \simeq 1.2 \times 10^{-10} {\rm m}\,{\rm s}^{-2}$, and in
the present work we adopt $\mu(y)=y/\sqrt{1+y^2}$ (Milgrom~1983).
From the Poisson equation and equation~(\ref{eqMOND}) it follows that
the MOND ($\gv=-\nabla\phi$) and Newtonian ($\gvN=-\nabla \phiN$)
gravitational fields are related by ${\mu}(g/\az) \, \gv = \gvN +\Sv$,
where $g\equiv\Vert\gv\Vert$, and $\Sv$ is a solenoidal field
dependent on the specific $\rho$ considered: in general one cannot
impose $\Sv=0$, thus the use of standard Poisson solvers to develop
MOND N-body codes is not possible, and equation~(\ref{eqMOND}) must be
solved at each time step (Brada \& Milgrom~1999; NLC07).

\subsection{Initial conditions and the code}
\label{secini}
The baryonic component of the initial conditions of all the
simulations presented in this paper consists of two identical galaxy
models with stellar density distribution
\begin{equation}
\rhos(r)={\Mstar\over 2\pi}{\rs\over r (r+\rs)^3},
\label{eqrhostar}
\end{equation}
where $\Mstar$ is the total stellar mass and $\rs$ is the core radius
(Hernquist~1990). 
To each MOND model with potential $\phi$ corresponds an {\it
equivalent} Newtonian model with $\phiN=\phi$, thus having a DM halo
with density $\rhodm(r)=\nabla^2\phi(r) /4\pi G-\rhos(r)$. In
principle, such a DM halo would have infinite mass, so we truncate it
at $r\sim30\rs$.  For completeness, we also ran simulations of
Newtonian merging between purely baryonic systems with the same
stellar density distribution~(\ref{eqrhostar}) and no DM halo.

The particles of the stellar component are distributed with the
standard rejection technique applied to the phase-space distribution
function (DF), restricting for simplicity to the fully isotropic case.
In the purely baryonic Newtonian case the DF is known explicitly
(Hernquist~1990), while in MOND the corresponding DF is obtained
numerically with an Eddington inversion (e.g. Binney \& Tremaine~1987)
\begin{equation}
f_M(E)={1\over \sqrt{8}\pi^2}{d\over dE}\int_E^{\infty}{d\rhos\over d\phi}
{d\phi\over\sqrt{\phi -E}},
\label{eqdf}
\end{equation}
where the upper integration limit reflects the far-field logarithmic
behaviour of the MOND potential (see also Angus, Famaey \&
Zhao~2006).  Finally, in the equivalent Newtonian models the stellar
particles are distributed by using their numerical two-component
isotropic DF. However, this is not possible in general for the
equivalent DM halo particles, because for systems with sufficiently
high stellar surface density the resulting halo presents a central
hole, and so it cannot be derived from an everywhere positive,
isotropic DF\footnote{This result shows that it is important to check
the positivity of the DF (and not only that of $\rhodm$), when
studying Newtonian systems with DM equivalent to MOND models.}
(Ciotti \& Pellegrini~1992). Thus, the initial DM particle velocities
are extracted from a Maxwellian distribution with local velocity
dispersion satisfying the isotropic two-component Jeans' equations. We
verified that the resulting models are in approximate equilibrium by
evolving them in isolation for several dynamical times.

We consider both head-on and off-centre encounters. In the head-on
cases (impact parameter $\bz=0$) the two galaxies are released at
$t=0$ with barycentric distance $\dz=40\rs$, and with the relative
speed $\vz$ that they would have if they started at rest at
$\drest=60\rs$. Thus, in the Newtonian cases
\begin{equation}
\vz^2=4G(\Mstar+\MDM)\left({1\over\dz}-{1\over\drest}\right),
\label{eqvznewt}
\end{equation}
while in the MOND cases
\begin{equation}
\vz^2\simeq0.8\sqrt{8G\Mstar\az}\ln{\drest\over\dz},
\label{eqvzMOND}
\end{equation} 
where we have used the approximate expression of the force between two
particles in deep-MOND regime (Milgrom~1986; Milgrom~1994).  In the
off-centre cases, $\dz$ and $\vz$ are the same as in the corresponding
head-on cases, but the relative velocity is oriented so that the
impact parameter $\bz=\dz/2$. 

The physical scales of the problem are introduced as follows.  First
of all, we identify each MOND initial condition by fixing a value for
the dimensionless internal acceleration parameter $\kappa\equiv
G\Mstar/(\az\rs^2)$, so $\Mstar$ and $\rs$ are not independent
quantities: in physical units, $\rs\simeq3.4
\kappa^{-1/2}\Mstarten^{1/2}\kpc$, where $\Mstarten\equiv
\Mstar/10^{10}\Msun$. The time and velocity units are
$\tstar=\sqrt{\rs^3/G\Mstar} \simeq
29.7\kappa^{-3/4}\Mstarten^{1/4}\Myr$, and $\vstar = \rs/\tstar \simeq
112\kappa^{1/4}\Mstarten^{1/4}\kms$ (see NLC07 for a more detailed
discussion of the normalisations).  The simulations are evolved up to
$t=400\tstar$ ($\kappa=1$ cases) or $t=500\tstar$ ($\kappa=25$ and
purely baryonic Newtonian cases), which amount to several gigayears in
physical units for galaxy masses in the observed range.

\begin{figure}
\centerline{ \psfig{file=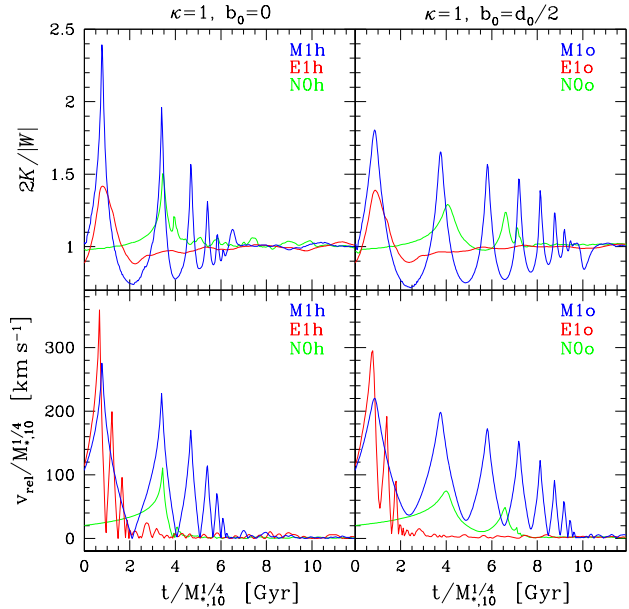,width=\hsize}}
\caption{Time evolution of the virial ratio (top panels) and of the
  barycentric relative speed (bottom panels) for the two sets of
  simulations with $\kappa=1$: head-on (left panels) and off-centre
  (right panels). Blue, red, and green curves refer to MOND,
  equivalent Newtonian, and purely baryonic Newtonian simulations,
  respectively.  For the scaling of time and velocity, see
  Section~\protect\ref{secini}.}
\label{figtime1}
\end{figure}

Our MOND N-body code (NLC07) is a parallel three-dimensional
particle-mesh code that can be used to run MOND as well as Newtonian
simulations. The code is based on a grid in spherical coordinates, on
which the MOND potential is computed by solving exactly the field
equation~(\ref{eqMOND}) with the iterative potential solver (based on
spectral methods) described in Ciotti et al.~(2006).  Particle-mesh
interpolations are obtained with a quadratic spline in each
coordinate, while time stepping is given by a classical leap-frog
scheme. The time step is the same for all particles and is allowed to
vary adaptively in time. Given the spherical geometry of the grid, the
code is not is not best-suited to run merging simulations: in order to
obviate this difficulty, we used a time-adaptive grid with a much
larger number of grid points ($128^3$) than in the collapse
simulations of NLC07. With this resolution, we obtained excellent
agreement between Newtonian merging simulations run with the MOND code
and simulations (starting from the same initial conditions) run with
our FVFPS treecode (Fortran Version of a Fast Poisson Solver;
Londrillo, Nipoti \& Ciotti~2003; Nipoti, Londrillo \&
Ciotti~2003). In summary, all the presented simulations were run with
the MOND code, and the Newtonian simulations also with the FVFPS code.
The properties of the simulations, including the total number of
stellar ($\Nstar$) and DM ($\NDM$) particles used, are summarised in
Table~1.


\begin{figure}
\centerline{
\psfig{file=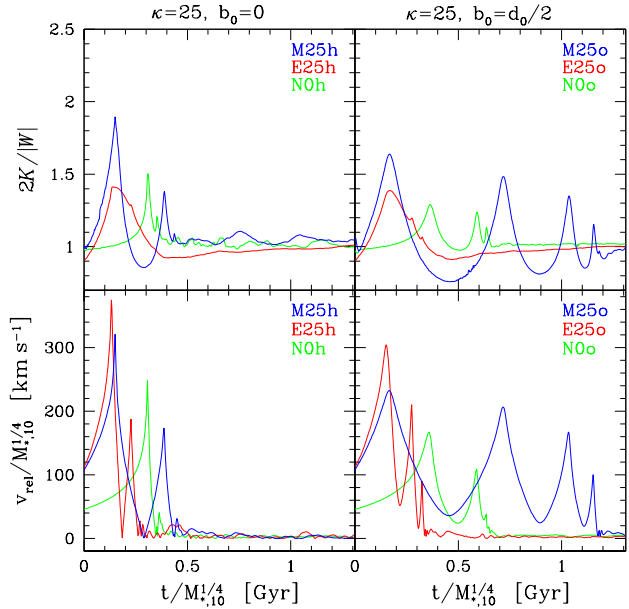,width=\hsize}}
\caption{The same as Fig.~\protect\ref{figtime1}, but for the two sets
of simulations with $\kappa=25$: head-on (left panels) and off-centre
(right panels).}
\label{figtime25}
\end{figure}

\section{Results}

\subsection{Merging dynamics and timescales}
\label{sectim}

We present now the results of four sets of merging simulations, each
characterised by a combination of the value of the internal
acceleration parameter ($\kappa= 1$ or $\kappa=25$) and of the impact
parameter ($\bz=0$ or $\bz=\dz/2$).  Each set comprises three
simulations: MOND, equivalent Newtonian ($\MDM=30\Mstar$ for
$\kappa=1$, and $\MDM=5\Mstar$ for $\kappa=25$) and purely baryonic
Newtonian. In the top panels of Figs.~\ref{figtime1} and
\ref{figtime25} we show the time evolution of the virial ratio
$2K/|W|$ (where $K$ is the total kinetic energy and $W$ is the trace
of the Chandrasekhar potential energy tensor) and, in the bottom
panels, the time evolution of the relative speed $\vrel$ of the
barycentres of the two galaxies: note that the time and velocity units
are the same for all simulations.  Peaks in $2K/|W|$ and in $\vrel$
correspond to close encounters between the two systems, while minima
of $2K/|W|$ and $\vrel$ occur when the separation is maximum.  At the
end of all the presented simulations $2K/|W|\sim 1$ and $\vrel\sim0$,
indicating that the two systems merged, forming a single virialised
object.

Let us focus first on the case $\kappa=1$, in which the initial
galaxies, having internal accelerations everywhere lower than $\az$,
are in deep-MOND regime.  As can be seen from Fig.~\ref{figtime1}, in
both the head-on and the off-centre cases, the merging timescale is
apparently longer in MOND (blue curves) than in the equivalent
Newtonian simulations (red curves). In MOND the two galaxies
experience several close encounters before merging, while in the
equivalent Newtonian cases they merge quickly after the first close
passage. The behaviour of both MOND and purely baryonic Newtonian
cases (green curves) is very sensitive to whether the orbit is head-on
or off-centre: the merging timescale in simulation N0o is almost a
factor of two longer than in simulation N0h, and also simulation M1o
takes significantly longer to virialise than simulation M1h.  In
contrast, due to the presence of DM halos, the merging timescale is as
short in the off-centre as in head-on the equivalent Newtonian cases.
As expected, the relative speed during the first close encounter is
significantly higher in MOND simulations than in the purely baryonic
ones. On the other hand, the equivalent Newtonian models collide at
higher speed than their MOND counterparts. We note that this last
result depends on the specific choice of $\MDM$ and $\drest$ appearing
in equations~(\ref{eqvznewt}) and (\ref{eqvzMOND}): provided that
$\dz$ is large enough MOND would have no problem in attaining
arbitrarily high collision speeds (see also Angus \& McGaugh~2007 for
a discussion of the collision speed of galaxy clusters in MOND).

The case $\kappa=25$ (Fig.~\ref{figtime25}), in which the initial
galaxy models have internal accelerations $\gsim\az$ for $r\lsim5\rs$,
confirms the same trend as the $\kappa=1$ case, with merging taking
longer in MOND than in equivalent Newtonian models (by a factor of
$\sim 2$ in the head-on case, and by a factor of $\sim 4$ in the
off-centre case).  The $\kappa=25$ simulations are interesting also
because they show how the merging process in MOND is very different
from that of purely baryonic Newtonian merging, even when the MOND
galaxies are internally in Newtonian regime.  This can be easily seen
in the head-on simulation M25h, in which the dynamics is almost
Newtonian when the two galaxies interpenetrate, but the collision
speed is higher than in the purely baryonic Newtonian case, being
determined by the long-range deep-MOND interaction between the two
galaxies.

We note that the value of $\kappa$ contains information only on how
the {\it initial} internal accelerations compare with $\az$. A model
initially characterised by accelerations everywhere weaker than $\az$
can produce accelerations significantly stronger than $\az$ during its
dynamical evolution.  This behaviour was observed by NLC07 in MOND
dissipationless collapse simulations (see also Nusser \&
Pointecouteau~2006, who studied spherically symmetric MOND gaseous
collapses in a cosmological context).  However, this is not
necessarily the case in MOND merging, because during the
dissipationless merging process the density does not increase as much
as in a collapse.  To quantify this effect we computed at each time
step the fraction of particles with acceleration stronger than
$\az$. In simulations M25h and M25o this fraction is initially
$\sim0.3$, has a peak up to $\sim0.5$ during the first close passage,
and is again $\sim 0.3$ in the end-products. On the other hand, it
turns out that in simulations M1h and M1o this fraction is $\lsim
0.02$ throughout the entire simulation: in other words our $\kappa=1$
simulations are in deep-MOND regime at all times.

From an observational point of view, simulations with $\kappa=1$ can
represent merging between two dwarf spheroidal galaxies with
$\Mstar=10^7\Msun$ (and effective radius $\Re \sim 0.2\kpc$, so the
merging timescale would be $\lsim 1.8 \Gyr$), while simulations with
$\kappa=25$ can represent merging between two luminous elliptical
galaxies with $\Mstar\sim10^{11}\Msun$ (and effective radius $\Re \sim
4\kpc$, so the merging timescale would be $\lsim 2.1 \Gyr$).  Thus,
restricting to the presented cases, one could be tempted to conclude
that galaxies in MOND can merge in a timescale significantly shorter
than the Hubble time.  However, we stress that such a conclusion is
wrong in a general sense, because we reported only cases with orbital
energies corresponding to two galaxies at rest when at relatively
small distance ($\drest\simeq25\rhalfz$, where $\rhalfz$ is the
half-mass radius of the initial stellar distributions).  Given the
logarithmic nature of the MOND gravitational potential (see
equation~\ref{eqvzMOND}), choosing larger values of $\drest$ has the
effect of boosting the encounter relative speed (making the merging
process difficult), while it barely affects the encounter speed in
Newtonian gravity.  In fact, we explored several other cases of MOND
encounters, with larger $\drest$ (and then higher $\vz$), but we had
to stop the simulations, because the two galaxies after the first
close passage reach relative distances significantly larger than
$\dz$, making the required computational time exceedingly long,
revealing virialisation times even longer than the age of the Universe
(note that the MOND simulation in the right panels of
Fig.~\ref{figtime1} is already dangerously long).  Summarising, we
presented here only simulations of encounters relatively favourable to
merging in MOND, and yet these mergings were found to be less
effective than in Newtonian gravity with DM.

\subsection{Merging end-products}

We define merging end-products the systems comprising the bound {\it
stellar} particles at the end of the simulation. In the Newtonian
simulations we found that $\lsim 4$ per cent of the stellar particles
escaped, while there cannot be escapers in the MOND cases. Using the
same procedure as in NLC07, we determined the following properties of
the end-products: the axis ratios $c/a$ and $b/a$ of the inertia
ellipsoid, the angle-averaged half-mass radius $\rhalf$, the virial
velocity dispersion $\sigmav$, the inner slope $\gamma$ of the
$\gamma$-model (Dehnen~1993; Tremaine et al.~1994) that best fits the
final angle-averaged density profile (over the radial range
$0.1\,\lsim\, r/\rhalf\,\lsim\, 10$), and, for the three principal
axis projections, the circularised effective radius $\Re$ and the
index $m$ of the Sersic~(1968) law that best fits the circularised
projected density profile over the radial range $0.1\,\lsim\,
R/\Re\,\lsim\, 10$ (see Table~1, where $\mav$ is the average of the
values of $m$ obtained for the three projections).

The structural and kinematic properties of the MOND end-products are
not significantly different from those of their Newtonian equivalent
counterparts: for instance, the final axis ratios are roughly the same
in corresponding MOND and equivalent Newtonian simulations (see
Table~1).  The end-products of simulations E1h and E1o are DM
dominated at all radii, and similarly the end-products of the
$\kappa=1$ MOND mergers are everywhere in deep-MOND regime (so they
would appear as DM dominated at all radii if interpreted in Newtonian
gravity). On the other hand, the $\kappa=25$ MOND end-products would
appear in Newtonian gravity as baryon-dominated in the inner regions
($r/\rhalf\lsim 1.2-1.3$) and DM dominated at larger radii; the
corresponding equivalent Newtonian end-products are baryon dominated
at radii $r/\rhalf \lsim 0.4-0.5$. In general both MOND and equivalent
end-products have rather flat intrinsic and projected velocity
dispersion profiles. The MOND final density profiles tend to be
steeper ($\gamma=1.4-2.0$, $\mav=4.2-6.3$) than those of the
equivalent Newtonian cases ($\gamma=1.0-1.7$, $\mav=3.4-4.5$), but
there is not a dichotomy between the two families.

An interesting point (in the context of the galaxy scaling relations)
is how the final virial velocity dispersion $\sigmav$ and half-mass
radius $\rhalf$ compare with the corresponding quantities in the
initial systems $\sigmavz$ and $\rhalfz$ (Nipoti et al.~2003).  MOND
mergers have larger $\rhalf$ and lower $\sigmav$ than the
corresponding equivalent Newtonian mergers.  We also note that in the
Newtonian cases here presented the ratio $\sigmav/\sigmavz$ tend to be
larger (and $\rhalf/\rhalfz$ smaller) than in similar cases explored
in Nipoti et al. (2003): this is expected, because here we consider
elliptic orbits while Nipoti et al.~(2003) considered parabolic
orbits.

\section{Discussion and conclusions}
\label{secdis}

The main result of the present work is that galaxy merging is much
less effective in MOND than in Newtonian dynamics with DM. In
addition, the derived MOND merging timescales must be considered only
lower limits, because rather specific orbital properties are required
in MOND in order to have galaxy mergers on timescales shorter than the
age of the Universe.  In general, repeated high speed galaxy
encounters should be a common feature of galaxy interactions in MOND,
while any observational evidence of rapid merging after the first
close passage should be regarded as an indication of the presence of
DM halos. Remarkably, when the orbital parameters are favourable and
two galaxies eventually merge in MOND, the merging end-product is
hardly distinguishable from the final stellar distribution of an
equivalent Newtonian merger with DM.

Thus, the very observation of galaxy mergers appears to favour the DM
scenario with respect to the MOND hypothesis. Additional constraints
for galaxy merging in MOND could be also given by specific dynamical
features in galaxy interactions that have extensively studied and
explained in the context of Newtonian gravity (e.g. Binney \&
Tremaine~1987), such as the tidal tails observed around interacting
disk galaxies as the ``Antennae'' pair of galaxies NGC\,4038/NGC\,4039
(Toomre \& Toomre~1972), and the surface brightness ripples observed
in the outskirts of luminous elliptical galaxies as NGC\,3923
(Quinn~1984).

The result that merging is less effective in MOND than in a DM
scenario appears consistent with our previous findings that
phase-mixing and violent relaxation are slower in MOND than in
Newtonian gravity (NLC07; Ciotti et al.~2007).  The merging process is
intimately related also to dynamical friction, so our simulations
might be interpreted as an indication that dynamical friction is less
effective in MOND than in Newtonian gravity with DM, in contrast with
the analytical estimates of Ciotti \& Binney (2004) for the case of a
particle moving in a homogeneous medium.  However, the complexity of
the merging process prevents us from drawing firm conclusions on this
issue, and we plan to realise ad hoc numerical experiments to explore
in detail dynamical friction in MOND.

We must also recall that we explored only very simple cases of galaxy
merging in MOND: in particular, we only considered equal-mass
dissipationless merging between spherical systems, while dissipative
processes in the merging of gas-rich galaxies might be effective in
making the merging timescales shorter.  Another possible caveat is
that, given the long-range nature of MOND gravity, the restriction to
an isolated pair of galaxies might not be as justified as in Newtonian
gravity, and the next step to address this point would be the study of
galaxy merging in MOND in a cosmological context. A valuable
contribution in this direction would be the performance of
cosmological simulations of structure formation, based on a
self-consistent relativistic formulation of MOND such as Bekenstein's
(2004) TeVeS.

\section*{Acknowledgements}

We are grateful to James Binney and Alar Toomre for helpful
discussions. We also thank the anonymous Referee for useful comments
on the manuscript. Some of the numerical simulations were performed
using the CLX system at CINECA, Bologna, with CPU time assigned under
the INAF-CINECA agreement 2006/2007.

\end{document}